\documentclass[12pt]{article}
\pdfoutput=1
\usepackage{graphics, color}
\usepackage{graphicx}
\usepackage{amsmath}
\usepackage{amssymb}
\usepackage{amsfonts}
\usepackage{tensor}
\usepackage[usenames,dvipsnames]{xcolor}
\usepackage[normalem]{ulem}
\usepackage[bottom]{footmisc}

\definecolor{orange}{rgb}{1,0.5,0}
\definecolor{grey}{rgb}{.5,.5,.5}
\definecolor{bluegreen}{rgb}{0,.5,.5}
\definecolor{darkgreen}{rgb}{0,.5,0}

\def\gsim{\, \rlap{$>$}{\lower 1.1ex\hbox{$\sim$}}\,}
\def\lsim{\, \rlap{$<$}{\lower 1.1ex\hbox{$\sim$}}\,}
\newcommand{\be}{\begin{equation}}
\newcommand{\ee}{\end{equation}}
\newcommand{\bea}{\begin{eqnarray}}
\newcommand{\eea}{\end{eqnarray}}

\newcommand{\mypar}{{\scriptscriptstyle \|}}

\newcommand{\overbar}[1]{\mkern 1.5mu\overline{\mkern-1.5mu#1\mkern-1.5mu}\mkern 1.5mu}

\begin{document}

%Title page

\begin{titlepage}
\bigskip
\bigskip\bigskip\bigskip
%\centerline{\Large \bf Something something something}
\centerline{\Large \bf Remarks on brane and antibrane dynamics}

\bigskip\bigskip\bigskip
\bigskip\bigskip\bigskip

 \centerline{{\bf Ben Michel,}\footnote{\tt michel@physics.ucsb.edu}
 {\bf Eric Mintun,}\footnote{\tt mintun@physics.ucsb.edu}
 {\bf Joseph Polchinski,}\footnote{\tt joep@kitp.ucsb.edu }${}^\dagger$}
 \centerline{ {\bf Andrea Puhm,}\footnote{\tt puhma@physics.ucsb.edu }
  and {\bf Philip Saad}\footnote{\tt phil.saad333@gmail.com }
 }
 \bigskip
\centerline{\em Department of Physics}
\centerline{\em University of California}
\centerline{\em Santa Barbara, CA 93106 USA}
\bigskip
\centerline{\em ${}^\dagger$Kavli Institute for Theoretical Physics}
\centerline{\em University of California}
\centerline{\em Santa Barbara, CA 93106-4030 USA}

\bigskip\bigskip\bigskip
%ABSTRACT

\begin{abstract}
We develop the point of view that brane actions should be understood in the context of effective field theory, and that this is the correct way to treat classical as well as loop divergences.  We illustrate this idea in a simple model.  We then consider the implications for the dynamics of antibranes in flux backgrounds, focusing on the simplest case of a single antibrane.  We argue that that effective field theory gives a valid description of the antibrane, and that there is no instability in this approximation.  Further, conservation laws exclude possible nonperturbative decays, aside from the well-known NS5-brane instanton.

\end{abstract}
\end{titlepage}

\baselineskip = 16pt
\tableofcontents

\baselineskip = 18pt

\setcounter{footnote}{0}

%\tableofcontents

\section{Introduction}

Brane actions are important for understanding many aspects of string physics.  However, their precise interpretation is somewhat ambiguous.  A brane is a source for the bulk fields, which are singular at the brane itself.  If these fields are then inserted into the brane action, the result is divergent.  Many applications use a probe approximation, in which the self-fields of a brane are not included in the brane's action.  This is like a formal limit in which the number of branes goes to zero.  

A more general approach is to interpret the brane action in the context of effective field theory.  Here, all effects are included, and divergences are treated via the usual framework of EFT~\cite{Weinberg:1978kz}.  For brane actions, this has been developed in Ref.~\cite{Goldberger:2001tn}, which shows that renormalization is the appropriate tool even for classical divergences such as those described above.\footnote{Related earlier work includes Refs.~\cite{Allen:1995rd}.} This can even lead to renormalization group flows of the type usually associated with quantum loops. 
In this paper we develop this point of view further, and show that it is useful in resolving some vexing issues in the literature.

In \S2 we present a simple model that illustrates how the framework of Ref.~\cite{Weinberg:1978kz} applies to branes.  We discuss the matching onto the UV theory in various cases.  In \S3 we apply the EFT point of view to anti-D-branes in a flux background, focusing primarily on the case of a single antibrane.\footnote{For a review
of the extensive literature on the supergravity descriptions of antibranes in flux backgrounds and a complete list of references see~\cite{DMVreview}.}  We recover the phenomenon~\cite{DeWolfe:2004qx,Blaback:2012nf} that in a flux background both branes and antibranes are screened by a background charge of the opposite sign.  Divergences of the screening cloud near the brane are resolved by matching onto string theory at short distance and are not sources of instability.  We show that possible nonperturbative annihilation of the antibrane and polarization cloud, while consistent with conservation of brane charge, is inconsistent with the $H_3$ Bianchi identity.  Further, the apparent impossibility of black branes with antibrane charge~\cite{Bena:2012ek,Bena:2013hr,Blaback:2014tfa} is avoided by proper account of a Bohm-Aharonov phase.  The only allowed antibrane instability is the NS5-brane instanton of Ref.~\cite{Kachru:2002gs}.

\section{Effective brane actions}

We illustrate the principle of effective brane actions with a simple model that captures the classical divergence problem noted above, and which gives a nice illustration of the general framework of Ref.~\cite{Weinberg:1978kz}.  In this model, the only bulk field is a free massless scalar $\phi$ in $d$ spacetime dimensions.  For now the brane is fixed on a $p+1$ dimensional subspace $x^{p+1} = \ldots = x^{d-1} = 0$, and it interacts with the bulk field via a general function of $\phi$ and its derivatives,
\be
S = -\frac{1}{2} \int d^dx \, \partial_M \phi \partial^M \phi + \int d^{p+1} x_\mypar \, {\cal L}_{\rm brane}(\phi,\partial) \,.  \label{act}
\ee
We will use $M,N$ for all $d$ dimensions, $\mu,\nu$ for directions tangent to the brane, and $m,n$ for directions orthogonal to the brane.
For given $d$ and $p$ there will be only a finite number of renormalizable interactions, but in the spirit of effective field theory we keep all interactions, with nonrenormalizable interactions suppressed by the appropriate power of a large mass scale $\Lambda$. We are imagining that the brane is described in a UV complete theory such as string theory, in which these general interactions will be generated.  If we are interested in amplitudes to some specified accuracy in $1/\Lambda$, then only a finite number of interactions contribute~\cite{Weinberg:1978kz}.

This point of view also requires that we keep general interactions in the bulk, but for simplicity we have omitted these.  The form~(\ref{act}) is stable under renormalization.  To make things even simpler, we restrict the brane action to terms quadratic in $\phi$, but with arbitrary derivatives.  Again, this form is stable under renormalization.

To begin, we consider the simple interaction $\frac12 g \phi^2$.  To first order, Fig.~1a, the amplitude for $k_1 \to k_2$ scattering in the presence of the brane is
\be
{\cal T}^{(1)} = g (2\pi)^{p+1} \delta^{p+1}(k_{1\mypar} - k_{2\mypar}) \equiv g \delta_\mypar  \,. \label{t1}
\ee
Only momenta parallel to the brane are conserved, and we abbreviate the ubiquitous $\delta$-function as indicated.
\begin{figure}[!ht]
\begin{center}
%\vspace {-5pt}
\includegraphics[width=5in]{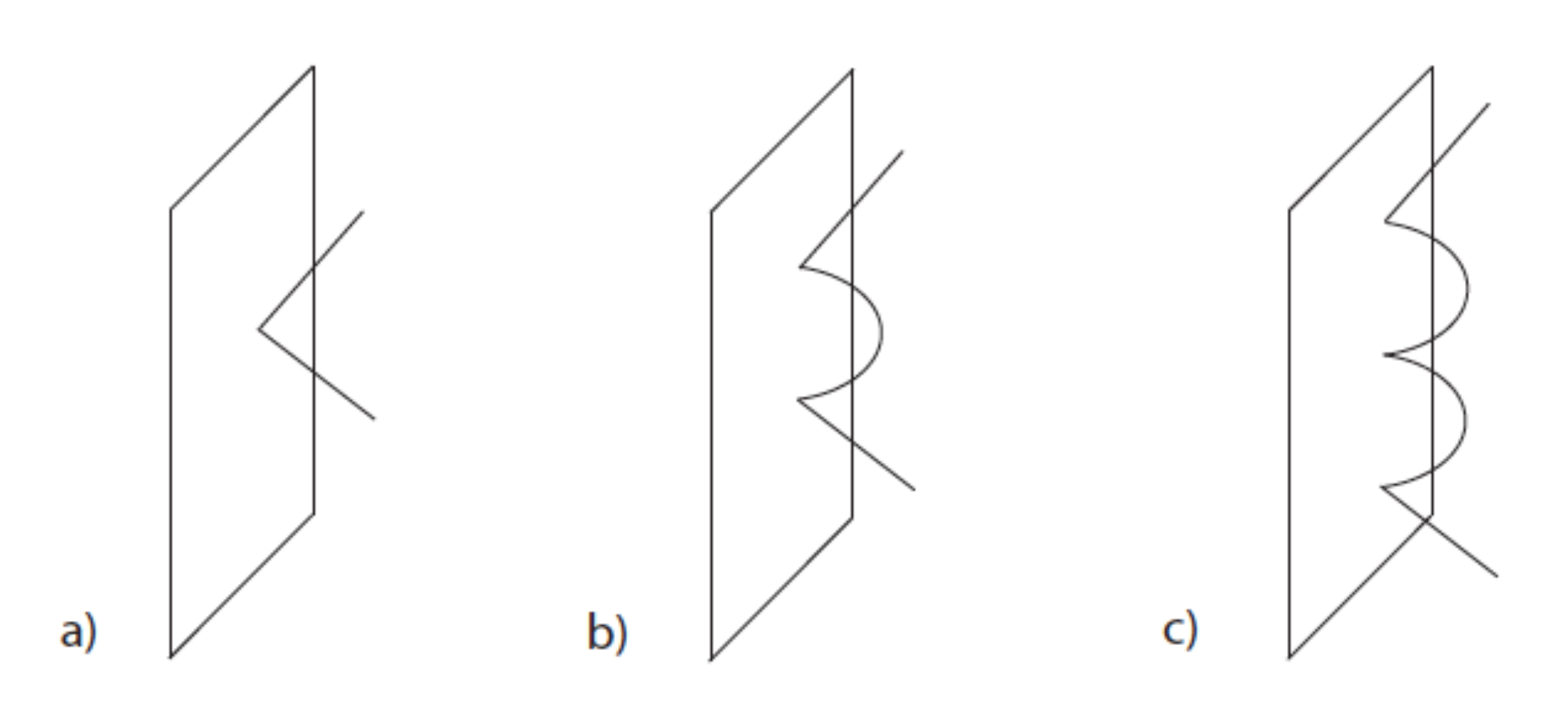}
\end{center}
%\vspace {-10pt}
\caption{First, second, and third order terms in the amplitude for $\phi$ to scatter from the brane.}
%\label{fig:radii}
\end{figure}

At second order, Fig.~1b, the amplitude is
\be
{\cal T}^{(2)} = g^2 \delta_\mypar   \int \frac{d^r k_\bot}{(2\pi)^r}
\frac{1}{k_\mypar^2 + k_\bot^2} \,.
\ee
Here $r = d-p-1$ is the number of transverse dimensions.  We see that this integral diverges for $r \geq 2$.  To analyze this, we cut the integral off at $k_\bot = \Lambda$, giving
\bea
\int ^\Lambda\frac{d^r k_\bot}{(2\pi)^r}
\frac{1}{k_\mypar^2 + k_\bot^2} &=& (-1)^n \pi C_r k_\mypar^{r-2} + 2 C_r \sum_{q=0}^\infty (-1)^q \frac{k_\mypar^{2q} \Lambda^{r - 2 - 2q}}{r-2-q} \,,\quad r=2n+1 \,,\nonumber\\
&=& (-1)^n C_r k_{ \mypar}^{r-2} \ln \frac{k_{ \mypar}^2}{\Lambda^2} + 2 C_r \sum_{q=0 \atop q \neq n-1}^\infty(-1)^q \frac{k_\mypar^{2q} \Lambda^{r - 2 - 2q}}{r-2-2q} \,,\quad r=2n
\,. \label{transint}
\eea
Here $C_r = V_{r-1}/2(2\pi)^r$ and $V_{r-1}$ is the volume of the unit $S^{r-1}$.
To analyze the divergences, let us note that the dimension of the interaction $\int d^{p+1}x \,\phi^2$ is 
\be
\Delta = d - p - 3 = r-2\,.
\ee
We include the volume element in the dimension, so negative $\Delta$ is relevant, vanishing $\Delta$ is marginal, and positive $\Delta$ is irrelevant (nonrenormalizable).

For codimension $r=1$, the integral converges.  Dropping for now terms suppressed by powers of $\Lambda$ (we will return to them later), we have
\be
{\cal T}^{(2)} = \frac{g^2  }{2  k_\mypar}\delta_\mypar  \,.
\ee
This dominates the leading term~(\ref{t1}) in the IR, as it should because the interaction is relevant.  Further graphs form a geometric series, beginning with Fig.~1c, giving in all
\be
{\cal T} = \frac{2 g  k_\mypar  }{2  k_\mypar - g} \delta_\mypar \,.
\ee
The interaction is attractive for positive $g$, consistent with the formation of a bound state.

For codimension $r=2$, there is a log divergence,
\be
{\cal T}^{(2)} = - \frac{g^2  }{4\pi} \delta_\mypar  \ln \frac{k_{ \mypar}^2}{\Lambda^2} \,.
\ee
Again we can sum the geometric series,
\be
{\cal T} = \frac{1 }{\frac{1}{g} + \frac{1}{4\pi}\ln \frac{k_{ \mypar}^2}{\Lambda^2}}\delta_\mypar  \,.
\ee
The appearance of a logarithm is not surprising because for $r=2$ the interaction is marginal.  These logarithms and their RG interpretation were discussed in Ref.~\cite{Goldberger:2001tn}.  
In conventional renormalization theory, we would take $\Lambda \to \infty$ holding fixed
$g(\mu)^{-1} = g^{-1} + \frac{1}{4\pi} \ln \mu^2/\Lambda^2$.
In effective field theory, $\Lambda$ is a fixed UV scale.  The divergence means that the effective field theory calculation is sensitive to UV physics, but only through local terms.   We need to adjust $g$ at this order, to account for the difference between our simple UV cutoff and the cutoff given by the true UV physics.  We will discuss the matching onto the UV theory below.
The logarithm means that the effective coupling $g(\mu)$ runs at scales below $\Lambda$.  For positive $g$ (attractive) there is again a pole in the IR, indicating a bound state.  For negative $g$ there is a Landau pole in the UV, but this is not a concern because this is only an effective theory.

For $r=3$ the story is similar but the divergence is linear.  The interaction is nonrenormalizable, so generically one would need more counterterms at higher loops, but in this simple model the higher loop graphs are just powers of the one loop graph and additional divergences do not appear.

For $r=4$ we have 
\be
{\cal T}^{(2)} = {g^2 C_4}\left(\Lambda^2 + k_{ \mypar}^2\ln \frac{k_{ \mypar}^2}{\Lambda^2}\right) \delta_\mypar \,.
\ee
Now there are quadratic and logarithmic divergences, so the result depends on two parameters from the UV theory.
The quadratic divergence requires adjustment of the original $g$ to match onto the short distance theory.  The log divergence requires a new interaction, $(\partial_\mypar \phi)^2$.  To make the power counting clearer we define a dimensionless coupling $\kappa_{00} = g \Lambda^{r-2}$, so that for $r=4$, 
\bea
{\cal T}^{(1)} &=& \frac{\kappa_{00}}{\Lambda^2} \delta_\mypar  \,, \nonumber\\
{\cal T}^{(2)} &=& {\kappa_{00}^2 C_4}\left(\frac{1}{\Lambda^2} + \frac{k_{ \mypar}^2}{\Lambda^4} \ln \frac{k_{ \mypar}^2}{\Lambda^2}\right) \delta_\mypar \,.
\eea
Because the interaction is irrelevant, $\Delta = 2$, even its leading effect is proportional to a negative power of $\Lambda$.   The second order $k_{\mypar}$-independent term is of the same order in $\Lambda$.  The $k_\mypar^2$ interaction comes with  $\Lambda^{-4}$, as appropriate for a $\Delta = 4$ interaction.  Its effect is suppressed relative to the $\Delta = 2$ term, but in the spirit of effective field theory we may be interested in $1/\Lambda$ corrections.  Note that the first nonanalyticity in $k_\mypar^2$ comes in at order $\Lambda^{-4}$. 

Note that all we are doing is solving the classical field equation
\be
\partial_{\mypar}^2 \phi + \partial_{\bot}^2 \phi = - g \delta^r(x_\bot) \phi \,,
\ee
%for the ${\cal L}_{\rm brane} = \frac12 g \phi^2$ interaction, 
but that this brings in the full machinery of EFT.
Note also that with $i \partial_t$ replaced by $\partial_{\mypar}^2$ this is the same as the Schrodinger equation with a $\delta$-function potential~\cite{Mintun:2014aka} (this has also been noted by the authors of Ref.~\cite{Goldberger:2001tn}).  The bound states that we have found for $r=1,2$ are well-known as bound states in the Schrodinger problem.  The case $r=2$ is often used as a simple model of renormalization.  Our discussion makes this connection more precise, and shows further that in general codimension this system also provides a simple model of EFT.  (The cases $r=2,3$ are discussed in Ref.~\cite{Jackiw:1991je}.)

In our toy example, where we have artificially restricted to interactions linear in $\phi^2$, the most general brane action would be
\be
S_{\rm brane} =  \frac{1}{2} \int d^{p+1} x_\mypar \, \sum_{l,j=0}^\infty \sum_m \,\frac{\kappa_{lj}}{\Lambda^{2l + 2j + r - 2}}
T^{jm}( \partial_\bot ) \partial_{\mu_1} \ldots \partial_{\mu_l} \phi\, T^{jm}( \partial_\bot ) \partial^{\mu_1} \ldots \partial^{\mu_l} \phi  \,.
\label{genact}
\ee
Here $T^{jm}$ is a traceless polynomial of degree $j$, and $m$ runs over these polynomials.  In writing this we have used field redefinition to remove terms containing $\partial_\bot^2$, and have integrated by parts with respect to $\partial_\mypar$ but not $\partial_\bot$.
To study amplitudes to accuracy~$\Lambda^{-s}$, one would retain all terms with $\Delta \leq s$.

We have omitted the brane's motion for simplicity, but this is readily included.  A simple model, in which we do not try to keep the full $d$-dimensional Lorentz invariance, adds in a transverse collective coordinate~$X^m(x_\mypar)$, beginning with the action
\bea
S &=& -\frac{1}{2} \int d^dx \, \partial_M \phi \partial^M \phi + \int d^{p+1} x_\mypar \left( -\frac{\tau}{2} \partial_\mu X^m \partial^\mu X^m
+ \frac{g}{2} \phi^2(x_\mypar,X_\bot(x_\mypar)) \right) \nonumber\\
&=& -\frac{1}{2} \int d^dx \, \partial_M \phi \partial^M \phi + \int d^{p+1} x_\mypar \left( -\frac{\tau}{2} \partial_\mu X^m \partial^\mu X^m
+ \frac{g}{2} \bigl[\phi(x_\mypar,0) + X^m \partial_m \phi(x_\mypar,0) + \ldots\bigr]^2  \right)\,. \nonumber\\
\eea
This now describes brane motion, and processes where scattering of scalars from the brane is accompanied by excitation of oscillations of the brane.  Locality and translation invariance imply that undifferentiated $X^m$'s appear only as arguments of $\phi$.  The same principles of renormalization apply.  These principles apply further for brane and bulk actions with general fields, including all interactions allowed by symmetry.

Note that in ${\cal T}^{(2)}$ the brane is interacting with its own induced field, so this goes beyond the probe approximation.  We have seen that this contribution is important for the leading IR physics for $r = 1,2$.  For larger $r$, it gives the leading nonanalytic behavior.  However, in many situations only the leading behavior in $1/\Lambda$ is of interest.  In particular, for branes of high codimension, the probe approximation ${\cal T}^{(1)}$ will be sufficient for most purposes.  The point of this exercise is just to illustrate that brane actions can be sensibly interpreted in the framework of effective field theory.  

Given a UV theory (we will consider some examples below), the couplings such as $\kappa_{lj}$ are determined by calculating some process  in both the UV theory and the effective theory with a given cutoff, and requiring that they agree.\footnote{The idea of matching is discussed in many reviews of effective field theory, e.g.~\cite{Manohar:1996cq}.}   After this is done, the effective theory can then be used for any other process.  Note that different cutoffs will give different values for the $\Lambda$-dependent terms in integrals such as~(\ref{transint}).  This is compensated by different values for the couplings in the effective theory.  It does not matter what cutoff we use as long as we are consistent, so in practice one often makes the simplest choice, dimensional regularization with minimal subtraction, for which
\bea
\int  \frac{d^r k_\bot}{(2\pi)^r}
\frac{1}{k_\mypar^2 + k_\bot^2} &=& (-1)^n \pi C_r k_\mypar^{r-2} \,,\quad r=2n+1 \,,\nonumber\\
&=& (-1)^n C_r k_{ \mypar}^{r-2} \ln \frac{k_{ \mypar}^2}{\mu^2}\,,\quad r=2n
\,. \label{transintdr}
\eea
The absence of power law divergences does not mean that the corresponding couplings are not generated: we still need to compensate for the difference between the dimensional regulator and the true UV physics.

Once the effective action is determined, it can be applied to other situations such as a brane in a background field $\overbar \phi$ (we use a bar to denote the background).  For example, the perturbation of the background by the brane is obtained from the same graphs as the $S$-matrix, in which one external state is replaced by $\overbar\phi$ and the other by a propagator, so the induced field is 
\be
\phi_{\rm ind}(k) = \frac{1}{k^2} \int \frac{d^d k'}{(2\pi)^d} {\cal T}(k,k') \overbar\phi(k') \,.
\ee
(Using ${\cal T}$ here is a slight abuse of notation, because $k'$ has been taken off-shell).  For illustration, using the probe approximation ${\cal T}^{(1)}$ with the general action~(\ref{genact}) gives in position space
\bea
\phi_{\rm ind}(x) &=&  \int d^{p+1} x'_\mypar \, \sum_{l,j,m} \,\frac{\kappa_{lj}}{\Lambda^{2l + 2j + r - 2}} T^{jm}( \partial'_\bot )  \partial'_{\mu_1} \ldots \partial'_{\mu_l} \overbar\phi(x')\times \nonumber\\
&&\qquad\qquad
T^{jm}( \partial_\bot ) \partial^{ \mu_1} \ldots \partial^{ \mu_l} \frac{1}{(d-2) V_{d-1}[(x_\mypar - x'_\mypar)^2 + x_\bot^2]^{(d-2)/2}}  \,.  \label{induced}
\eea
This diverges at the brane, and the divergence grows with $j$ and $l$.  However, the result applies only at momenta small compared to $\Lambda$ and so  at distances large compared to $\Lambda^{-1}$.  
Similarly, to study the motion of the brane in a background field, one can use the probe action or, if greater accuracy is needed, add in the higher corrections --- essentially $ {\cal T}$ again, but with both external states replaced by $\overbar\phi$ (we will see an example of this in Fig.~2b).

For a D-brane, the UV theory is string theory.  The amplitude ${\cal T}^{(1)}$ is the effective description of the disk with two closed string vertex operators.  By calculating this disk amplitude, and requiring that the effective field theory give the same answer, one determines the brane couplings with any number of derivatives (the equivalent to the $\kappa_{jl}$) to leading order in $g_{\rm s}$.\footnote{A partial list of papers on the disk effective action is given in Ref.~\cite{Gubser:1996wt}.}
 The amplitude ${\cal T}^{(2)}$ is the effective description of the annulus with two closed string vertex operators, and so one would need to match this amplitude to determine the effective action to order $g_{\rm s}^2$.

Another situation would be a solitonic brane, such as a magnetic monopole, vortex, or domain wall in a spontaneously broken \mbox{QFT}.  The UV theory would be the unbroken QFT and the effective theory would describe the brane collective coordinates plus any light fields.  Again one matches a UV calculation to one in the effective description.  In the UV calculation, the key input is the requirement that the fields of the soliton be nonsingular.  

One might try to apply the second method to the D-brane, using its supergravity description together with a condition such as~\cite{Gubser:2000nd} on allowable singularities.  However, the scale of the supergravity solution for a single D-brane is smaller than the string length by a power of $g_{\rm s}$, so this is not a good description.  (It is a valid description if enough D-branes coincide, a point we will return to below.)   If the supergravity approach did give an answer, it would likely not agree with the correct string theory result, because string theory knows about the scale $\alpha'$ and supergravity does not.  Similarly, if the supergravity approach fails to give an answer due to  singularities deemed bad, this has no physical significance.  It is the matching onto string perturbation theory that is the correct criterion for a good singularity in the fields external to a D-brane.\footnote{For M-branes there is no perturbative description of the UV theory, but Matrix theory~\cite{Banks:1996vh} 
provides a construction of the S-matrix to which the effective theory should be matched.}

For sufficiently supersymmetric amplitudes, the supergravity calculation will agree with the string calculation, because of the absence of $\alpha'$ corrections.  This does not mean that supergravity is an accurate description of a single D-brane. The magic of supersymmetry sometimes leads to complacency about the validity of effective descriptions.  For example, it has sometimes led to weak/strong dualities being misunderstood as weak/weak dualities.  

\section{Antibranes in fluxes}

\subsection{Application of EFT}

De Sitter vacua of string theory may be numerous but they are not simple.  (Meta)stability requires the balance of several forms of energy density~\cite{Silverstein:2001xn}.  The KKLT construction~\cite{Kachru:2003aw} begins with a  supersymmetric anti-de Sitter vacuum and excites it by adding one or more antibranes (branes having opposite supersymmetry to the background).  The nature of this supersymmetry breaking has recently been understood in Ref.~\cite{Kallosh:2014wsa}.  A body of work beginning with Refs.~\cite{McGuirk:2009xx,Bena:2009xk} has argued that the dynamics of anti-D-branes is complicated and potentially unstable. 

In the KKLT model~\cite{Kachru:2003aw}, a single antibrane can be sufficient to uplift an AdS vacuum to a dS vacuum, and this is the case that we focus on here.   The scale of the geometry is large compared to the string length, so EFT should be valid.  In the effective description the only low energy brane degrees of freedom are the gauge fields in the Poincar\'e directions and the collective coordinate for the brane motion.  The only thing the antibrane can do to lower its energy is to move to the position of lowest potential, the bottom of the Klebanov-Strassler (KS) throat.\footnote{Of course, if there are massless or light moduli in the vacuum without an anti-branes, adding the antibrane could destabilize them.  See for example Ref.~\cite{Dymarsky:2013tna}.  This would be seen in the effective field theory.}

To illustrate the use of EFT, consider a potentially problematic issue, the backreaction on $H_3$.  A low-order contribution is shown in Fig.~2a.  
\begin{figure}[!ht]
\begin{center}
%\vspace {-5pt}
\includegraphics[width=4in]{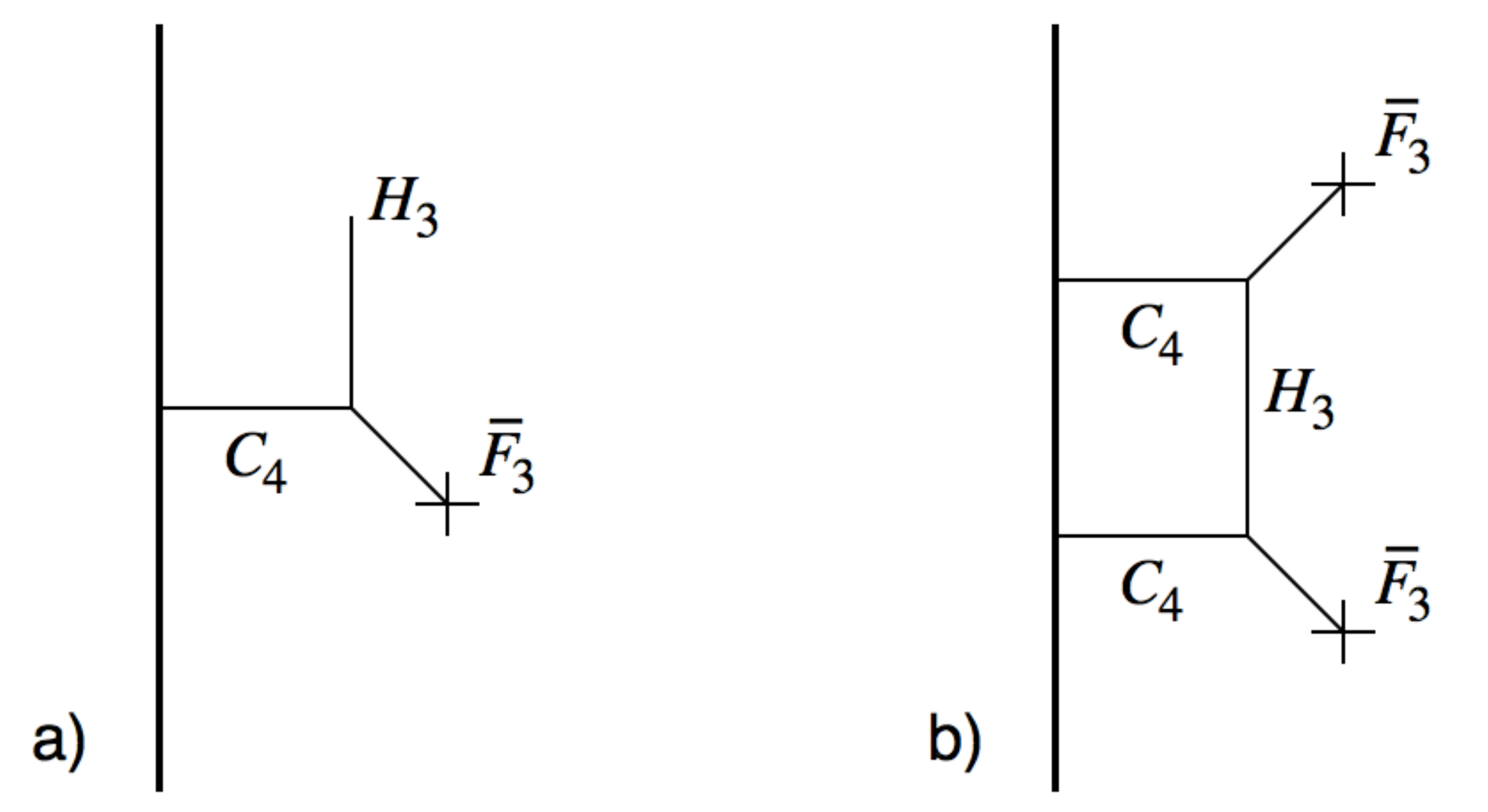}
\end{center}
\vspace {-10pt}
\caption{a) Lowest order backreaction on $H_3$.  The heavy line is the anti-D3 brane, and the $\times$ denotes a background field.   b) Corresponding contribution to the brane potential.}
%\label{fig:radii}
\end{figure}
A bulk potential has scaling dimension~4.  Its {\it engineering} dimension is 0 since we include an $\alpha'^{-4}$ in its kinetic term, but this is not what matters for degree of divergence; henceforth `dimension' refers to scaling unless otherwise specified.
The interaction $\alpha'^{-2} \int d^4 x \, C_4$ has  dimension $\Delta = 4-4 = 0$.  The Chern-Simons interaction $\alpha'^{-4} \int d^{10} x\, C_4 \wedge F_3 \wedge H_3$ has  dimension $\Delta = 4 + 5 + 5 - 10 = 4$.  The total dimension of all interactions in Fig.~2a is then $4$, and it follows that
\be
H_3 \propto \frac{g_{\rm s}^2 \alpha'^2 \overbar F_{3}}{x_\perp^4} \,, \label{hf}
\ee
where a bar again denotes the background field.  The $x_\perp^{-4}$ is from the scaling dimension, and the $\alpha'^2$ has been inserted by engineering dimensional analysis.  We work in the string metric, so the $g_{\rm s}^{2}$ is from the $H_3$ propagator.\footnote{The metric and $B_2$ have a $g_{\rm s}^{2}$ in the propagator; while the RR forms do not depend on $g_{\rm s}$.  The bulk gravitational interaction contains a $g_{\rm s}^{-2}$, while the Chern-Simons terms do not depend on $g_{\rm s}$.  The coupling of the metric to the brane is proportional to  $g_{\rm s}^{-1}$, while the coupling of the RR form to the brane does  not depend on $g_{\rm s}$.} There is also a contribution 
\be
H_3 \propto \frac{g_{\rm s} \alpha'^2 \overbar H_{3}}{x_\perp^4} \,, \label{hh}
\ee
from a similar graph with $g_{\mu\nu}$ in place of $F_5$ and $\overbar H_3$ in place of $\overbar F_3$. Even at the limit of EFT, $x_\perp \sim \alpha'^{1/2}$, this is a small perturbation on the background at weak coupling.

The integral of the energy density $ g_{\rm s}^{-2} H_3^2$ diverges quadratically at the brane.  The corresponding graph is Fig.~2b.  The total $\Delta$ of the interactions is $8$, and the leading brane counterterm $\int d^4 x\, F_3^2$ has dimension 6, so the divergence again comes out quadratic.  The counterterm is of order
\be
g_{\rm s}^{2} \alpha'^{-1} \int d^4 x\, \sqrt{-g_4} F_3^2 \,. \label{f3counter}
\ee
The factors of $\alpha'$ from the vertices and propagators cancel, leaving an $\alpha'^{-1} $ from the cutoff; the net result is fixed anyway by the (engineering) dimensions.  Again, the numerical coefficient would come from matching to the string annulus graph.  In a similar way we get a counterterm
\be
\alpha'^{-1} \int d^4 x\, \sqrt{-g_4} H_3^2 \,.  \label{h3counter}
\ee
The counterterms~(\ref{f3counter}, \ref{h3counter}) are each one order in $g_s$ higher than terms that are expected in the tree level brane action,
\be
g_{\rm s} \alpha'^{-1} \int d^4 x\, \sqrt{-g_4} F_3^2 \,, \quad 
g_{\rm s}^{-1}  \alpha'^{-1} \int d^4 x\, \sqrt{-g_4} H_3^2 \,,  \label{higherd}
\ee
as expected for the annulus in comparison to the disk.

Expanding around a minimum of the potential, one gets a mass correction of order
\be
\alpha'^{-1}  \left(g_{\rm s}^{2} (\partial \overbar F_3)^2 + (\partial \overbar H_3)^2 \right)X^2  \label{mcorr}
\ee
from the annulus corrections~(\ref{f3counter}, \ref{h3counter}).
Note that this is a dimensional estimate; the signs and tensor structures are not specified.  In particular, the  potential will vanish along Goldstone directions such at the $S^3$ of the Klebanov-Strassler throat.
For comparison, the leading order potential is $\alpha'^{-2} g_{\rm s}^{-1} \int d^4 x\, \sqrt{-g_4}$.  We can estimate the second derivative of this from Einstein's equation, giving a mass term of order
\be
\alpha'^{-2}  \left(g_{\rm s}  \overbar F_3^2 + g_{\rm s}^{-1}  \overbar H_3^2 \right)X^2 \,.
\ee
The mass correction~(\ref{mcorr}) is suppressed by $g_{\rm s}$ and also by $\alpha'/L^2$, where $L$ is the characteristic scale of the geometry.  The effect of the higher derivative tree-level terms~(\ref{higherd}) is suppressed by $\alpha'/L^2$ but not by $g_{\rm s}$.

In summary, self-consistent use of effective field theory shows no large corrections that would signal a breakdown.  Again, the antibrane's only degree of freedom is its position.  Energetically this is limited to a bounded space, the neighborhood of the bottom of the Klebanov-Strassler throat, and so there must be a minimum, where all perturbations have nonnegative mass-squared.

\subsection{More on antibrane dynamics}

When an antibrane and brane are close together, there is an open string tachyon between them that leads to their annihilation.  However, when the brane dissolves into flux, its world-volume gauge field is in a confining phase, and strings cannot terminate in flux, at least perturbatively.  There are no degrees of freedom within the EFT that would describe such an annihilation.  But the EFT does describe the dynamics of the fluxes, and a closer look at these is warranted.  

The antibrane can decay via an NS5-brane instanton~\cite{Kachru:2002gs}, which mediates the process 
\be
{\rm \overbar{D3}} + M {\ \rm units\ of\ }H_3 \wedge F_3 \ \to \ M-1 {\ \rm D3's} \,.
\ee
This is a nonperturbative effect.  The backreaction in effective field theory does not significantly affect the instanton action: the amplitude of Fig.\ 1c, for example, is further reduced by the dissolving of the $\overbar{\rm D3}$ in the NS5.  In particular, the flux-clumping Ansatz of Refs.~\cite{Blaback:2012nf,Bena:2013hr,Danielsson:2014yga} does not seem to apply.   

In the NS5 process~\cite{Kachru:2002gs}, the initial configuration is a stack of anti-D3-branes polarized into an NS5-brane that subtends an angle $\psi = \psi_i$.  The decay process involves $\psi$ tunneling through a potential barrier to a lower-energy final state.  For a single antibrane, the initial $\psi_i$ would be so small that the description breaks down: the initial polarization is negligible.  However, the decay process for a single antibrane still requires an NS5-brane instanton in order to source the $H_3$ Bianchi identity, so $\psi$ must pass through large values where the polarization picture applies, and the dominant contribution to the tunneling action comes from this region.  So the KPV result still applies for the single antibrane.
 
In Ref.~\cite{Bena:2012ek,Bena:2013hr}, it is shown that there is no black antibrane solution with $\overbar{\rm D6}$ charge immersed in a background of the opposite sign.\footnote{At zero temperature, if the antibrane $\delta$-function in Fig.~3b is smoothed, the density has a volcano shape with a maximum at the rim. Such a maximum at positive polarization density is forbidden~\cite{Blaback:2011nz} by Eqs.~(\ref{e1}-\ref{e3}), but this is at the string length and so outside the validity of effective field theory; there is no problem with the distribution in Fig.~3b in string perturbation theory.  A similar argument is used in the black case, but only outside the horizon where it is valid if the Schwarzschild radius is greater than the string length.}  This suggests that finite temperature eliminates the barrier to brane-flux annihilation so that it is rapid, rather than proceeding via tunneling.  However, even if this is true, it does not provide any evidence for rapid decay at zero temperature.    It is quite possible for a process to be nonperturbatively slow at low temperature and rapid at high temperature.  Electroweak baryon number violation is an important example.

In an earlier version of this work we suggested a more rapid, but still nonperturbative, decay.  In fact, this does not exist.\footnote{We thank Eva Silverstein and Juan Maldacena for pointing out our error.}  The remainder of this subsection deals with this.
We will focus on the anti-D6 case, which has been worked out in greatest detail~\cite{Blaback:2011nz,Bena:2013hr}.  The key field equations are
\bea
d(*_{10} e^{- 2 \phi} H_3) &=& -  F_0 {*}_{10} F_2 \,,  \label{e1} \\
dF_2 &=&  F_0 H_3 + \delta_{\rm D6} \,, \label{e2} \\
dF_{0} &=& 0 \label{e3} \\
dH_{3} &=& \delta_{\rm NS5}  \label{e4} 
\,.
\eea
We have translated to string frame for consistency with our earlier discussion.  For completeness every equation should include potential brane sources, but the F1 sources in~(\ref{e1}) and the D8 sources in~(\ref{e3}) will play no role; thus the zero form $F_0$ is constant.  The $\delta_{\rm D6}$ is summed with sign over brane and antibrane sources, and if the space transverse to the D6-branes is compact, it should also include negative contributions from O6 planes.

Expanding around this background we have in particular
\be
d\delta F_2 = F_0 \delta H_3 + \delta\delta_{\rm D6} \,. \label{p2}
\ee
The brane induces a $\delta F_2$ via Eq.~(\ref{e2}), and Eq.~(\ref{e1}) then leads to a $\delta H_3$.  On the RHS of Eq.~(\ref{p2}) this provides a perturbation to the background D6 density due to polarization of the flux background.  
Eqs.~(\ref{e1}, \ref{e2}) together imply a mass-squared term of order $e^{2\overbar \phi} \overbar F_0^2 \equiv \mu^2$ for the perturbations; essentially $F_2$ Higges $H_3$.  The perturbations thus fall exponentially away from the brane.    Integrating Eq.~(\ref{p2}) over the transverse space, the LHS must then vanish, and therefore the RHS does as well: the polarization of the background screens that of the D6 or anti-D6 completely~\cite{DeWolfe:2004qx,Blaback:2012nf}, Fig.~3.
\begin{figure}[!ht]
\begin{center}
\vspace {-5pt}
\includegraphics[width=6 in]{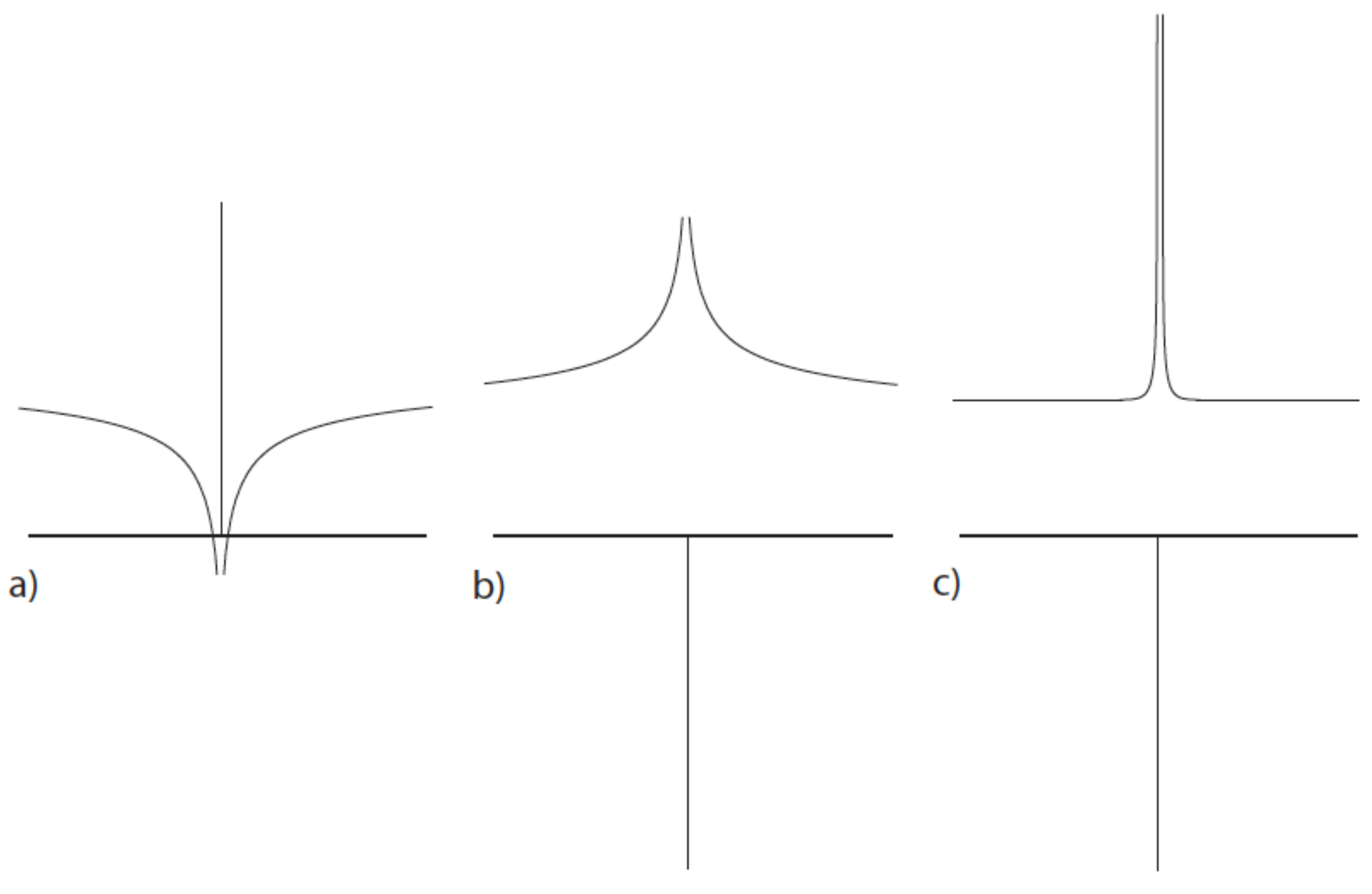}
\end{center}
\vspace {-10pt}
\caption{D6 densities in a flux background.  In all cases the excess or deficit in the screening cloud offsets that due to the brane source.
a) D6-brane in a flux background.  b) Anti-D6-brane in a flux background.  c) Fluctuation of the anti-D6-brane's screening cloud down to a size of order the string length.}
%\label{fig:radii}
\end{figure}

The total background charge contained within the volume of the screening cloud is of order 
\be
{\overbar H_3 \overbar F_0 }/{\mu^3} \sim {  e^{\overbar \phi} \overbar F_0^2}/{\mu^3} \sim {1}/{e^{2 \overbar\phi} \overbar  F_0} \,.
\ee
This is large if the flux is dilute and/or the coupling is weak, so we can treat the screening due to a single D6 as a perturbation.  
The screening cloud diverges as we approach the brane, and due to the nonlinearities of the field equation the expansion of the field near the origin will contain all negative powers of the distance.  However, as in the toy model example, this is not a problem: the brane effective action gives a precise prescription for matching the fields external to the brane onto the UV physics at the string length.   Further, we have seen from Eqs.~(\ref{hf}, \ref{hh}) that even very close to the brane the screening charge density is small, so cannot drive any open string model tachyon.

This discussion suggests that the antibrane can annihilate with its polarization cloud, ${\rm \overbar{D3}} + 1 {\ \rm units\ of\ }H_3 \wedge F_3 \to $ energy.  This was suggested in Refs.~\cite{Blaback:2012nf,Danielsson:2014yga} as a means of enhancing the NS5 decay; we have argued above that any such effect is slight.  The process considered in the earlier version of the present work was a local annihilation, ${\rm \overbar{D3}} + 1 {\ \rm unit\ of\ }H_3 \wedge F_3  \to  {\rm energy}$.  We will see that this is forbidden by the $H_3$ Bianchi identity.\footnote{
In the earlier version, the brane-flux annihilation was linked to the breaking of the heterotic string~\cite{Polchinski:2005bg}.  That process is consistent with the heterotic string Bianchi and quantization conditions, as shown explicitly there by a K theory construction.}

One might have thought that something interesting could happen nonperturbatively.  Consider a fluctuation of the supergravity fields like that shown in Fig.~3c, where the screening charge concentrates into a very small volume.  Is it possible for the brane and the flux to mutually annihilate?  This would conserve $D6$ charge, but we must also consider the $H_3$ Bianchi identity, which we can think of as a conservation law for the current $(*H)_7$.  The unit D6 charge of the polarization cloud implies $\int_{\rm cloud} H_3 = 1/F_0$, or 
\be
H_3 \sim \delta_{\overbar {D6}}/F_0
\ee
before the annihilation.  After the annihilation the cloud is gone, so there is a source
\be
dH_3 \sim \delta(t)\delta_{\rm\overbar {D6}}/F_0 \,.
\ee
This is geometrically consistent with an $NS5$-brane instanton, but that would give $dH_3 \sim \delta(t)\delta_{\rm\overbar {D6}}$.  The parametric dependence on $F_0$, which is an arbitrary integer in the natural units that we are using here, allows us to distinguish the known NS5-brane instanton process from the new brane-flux annihilation process suggested in version one.   We see that the latter is forbidden.

It is interesting to compare the zero-temperature and high-temperature behaviors.
In the absence of NS5-brane sources $H_3 = dB_2$.  Integrating the $F_2$ Bianchi identity~(\ref{e2}) on an $S^2$ just outside the black brane horizon, we get
\be
\frac{\partial}{\partial t} \int_{S^2} (F_2 - F_0 B_2) = 0 \,. \label{f2int}
\ee
We omit the source term because we will consider a process during which no branes cross the $S^2$.  The conserved quantity~(\ref{f2int}) was termed a Page charge in Ref.~\cite{Marolf:2000cb}.  As noted there, it is localized, quantized, conserved, but {\it not} invariant under large gauge transformations.  In particular it jumps by $F_0$ under $\int B_2 \to 1 + \int B_2$.  Thus it is a $Z_{F_0}$ charge, whose conservation excludes the brane-flux annihilation for a single antibrane.

Imagine starting with an antibrane at zero temperature.  In the integral~(\ref{f2int}), the flux from the antibrane contributes $-1$.  As the black hole forms, the argument of \cite{Bena:2012ek,Bena:2013hr} implies that it must absorb the polarization cloud in order that $ \int_{S^2} F_2$ becomes positive.  However, $\int_{S^2} (F_2 - F_0 B_2)$ remains negative and keeps track of the antibrane number.  If we cool the system back to zero temperature, the antibrane must reappear.  The integral of $B_2$ over the horizon is a sort of hair that can be measured in a stringy Bohm-Aharonov experiment~\cite{Bowick:1988xh}.  Again, their might be a process in which an NS5-brane instanton changes the Page charge by $F_0$ units, but it cannot change by a single unit.  Finite temperature would be expected to reduce the barrier for the NS5-brane instanton, and might eliminate it entirely at high enough temperature.\footnote{We thank Don Marolf for discussions of this point.}

More globally, imagine an $S^3$ whose equator $S^2$ surrounds the black 6-brane, and which is elongated in time to incorporate the black brane formation and disappearance.  Let the $S_3$ surround $N_{\rm NS5}$ NS5 instantons.  Then
\bea
\int_{S^3} H_3 &=& N_{\rm NS5} \,, \nonumber\\
0 = \int_{S^3} dF_2 &=& F_0   N_{\rm NS5} + \Delta N_{\rm D6} \,.
\eea
Thus the net $D6$ charge can only change in multiples of $F_0$.  The integer quantization of $\int_{S^3} H_3$ follows from the Dirac quantization condition~\cite{Nepomechie:1984wu}, independent of the dynamics internal to the $S^3$.

These considerations extend to other antibranes.\footnote{Gavin Hartnett has given a complementary argument for localized $\overbar{\rm D3}$'s, that there is no positivity condition on the black brane flux in this case~\cite{Hartnett:2015oda} .}   For the KKLT anti-D3, consider an $S^3_a \times S^3_b$, where $S^3_a$ is parallel to the bottom of the KS throat and $S^3_b$ surrounds the antibrane in the directions transverse to $S_a^3$, and in time.  Then 
\be
0 = \int_{S^3_a \times S^3_b} dF_5 = \int_{S^3_a \times S^3_b} (F_3 \wedge H_3 + \delta_{\rm D3}) = M N_{\rm NS5} + \Delta N_{\rm D3} \,.
\ee
Here $N_{\rm D3}$ includes any D3 charge arising from flux on wrapped 7-branes.  The $M$ is the number of units of $F_3$ flux on $S^3_a$. 
It plays the same role as $F_0$ above, distinguishing the known NS5 process~\cite{Kachru:2002gs} in which $\Delta N_{\rm D3} = -M$ from a potentially new brane-flux annihilation process with $\Delta N_{\rm D3} = -1$.  The latter is forbidden.   Thus the metastability estimates in the original work~\cite{Kachru:2003aw} appear to be correct.\footnote{The very recent work~\cite{Danielsson:2015eqa} has argued that the NS5 decay might be hastened by by passing through a new set of low energy configurations.  The proposed configurations violate the Bianchi identity for the NS5 world-volume gauge field, and so are forbidden.}

\subsection{Multiple antibranes}

For $p$ coincident D-branes, the effective field theory on the brane becomes non-Abelian (the notation $p$ is standard here, not to be confused with the dimension $p$ of the branes).  When $g_{\rm s} p \gg 1$, the brane theory is strongly coupled but the supergravity description is good.  For (anti-)D3-branes, the geometry near the branes is described by an $AdS_5 \times S^5$ throat at the bottom of the KS throat~\cite{Bena:2014jaa}.   When the background is slowly varying on the scale of the brane radius $(g_{\rm s} p)^{1/4} \alpha'^{1/2}$ (meaning that $p$ is parametrically smaller than $M$ in the KKLT context), one can again use an effective brane description of the system as seen from the outside.  In the UV, this is matched onto the supergravity description of the throat. Modes in the throat behave as $z^{\lambda_{\pm}}$.   Most modes correspond to irrelevant interactions, where $\lambda_- = - \Delta$ is negative and $\lambda_+ = \Delta + 4$ is positive (for consistency we continue to use the somewhat 
nonstandard convention that $\Delta$ includes $-4$ from the integration).  The 
$\lambda_-$ mode goes to zero at the bottom of the throat while the $\lambda_+$ mode diverges, and we get a good boundary condition by requiring that the latter vanish.  Integrating through the transition between the throat and the exterior to determine the exterior fields, and matching to an effective field theory calculation analogous to~(\ref{induced}), determines the parameters in the effective action.

However, for modes corresponding to relevant interactions, both $\lambda_+$ and $\lambda_-$ are positive and both modes grow down the $AdS_5$ throat.  In this case one must understand the nonlinearities there.  The \mbox{$\Delta=-1$} modes corresponding to fermion bilinears involve the 3-form fluxes.  For these the singularity is resolved by brane polarization~\cite{Polchinski:2000uf}, giving a good UV description.  The \mbox{$\Delta=-2$} modes corresponding to scalar bilinears involve the five-form flux and scalar deformation of the $S^5$.  Resolution of the resulting singularities requires the branes to move out onto the Coulomb branch~\cite{Warner:1999kh}, and because there is no $L=0$, $\Delta=-2$, scalar bilinear the potential is always negative in some directions and the (anti-)D3 branes are expelled by the AdS throat.  In either case, once the actual physics in the throat is understood, one can determine the effective field theory.
 
When both the $\Delta=-1$ and $\Delta=-2$ perturbations are present, there is a competition between these two effects.  
When the anti-D3-branes and their AdS throat are at the bottom of a KS throat, this is the case. This has recently been studied in Ref.~\cite{Bena:2014jaa}.  They concluded that if a  parameter Im$(\mu)$ is nonzero, then it is energetically favored for the branes to be expelled from the AdS throat.  They are expelled in an oblique direction (the so-called `giant tachyon' of~\cite{Bena:2014jaa}), so they are not precisely at the bottom of the KS throat, but energetically they cannot wander too far from the bottom.  The screening effect implies that antibranes attract at longer distances~\cite{DeWolfe:2004qx}, so their precise arrangement may be intricate, but in any case our earlier discussion of the single antibrane now applies.  
%The separated configuration is actually more stable than the polarized one, because the latter is already partway up the tunneling potential, so the bound $p < 0.08 M$~\cite{Kachru:2002gs} should be loosened.  
If the parameter Im$(\mu)$ vanishes (as may be required by symmetry), then the branes do polarize as in Ref.~\cite{Kachru:2002gs} if $p$ is not too large.  Again, this will be subject to nonperturbative decay via the NS5 instanton~\cite{Kachru:2002gs}.
 
\section{Conclusions}

We have argued that effective field theory allows the use of brane actions beyond the probe approximation, including the treatment of both classical and quantum divergences.  In all applications of brane systems, this provides a more general and physical interpretation of the results.  In applying this to the antibrane in flux, this validates the assumptions of Ref.~\cite{Kachru:2003aw}: the supersymmetry-breaking antibranes can be described by effective field theory, and  are metastable if their number $p$ is not too large.  It also follows that large classes of non-extremal fuzzball solutions (geometries or stringy solutions) using antibranes should exist~\cite{Bena:2012zi}.

\section*{Acknowledgments}

The work of B.M. is supported by the NSF Graduate Research Fellowship Grant DGE-1144085.  The work of E.M. is supported by NSF grant PHY13-16748.  The work of J.P. is supported by NSF Grant PHY11-25915 (academic year) and PHY13-16748 (summer).  The work of A.P. is supported by National Science Foundation Grant No. PHY12-05500.  The work of P.S. was supported by a Worster Fellowship.  We thank Iosif Bena, Ulf Danielsson, Mariana Gra\~na, Gavin Hartnett, Ehson Hatefi, Gary Horowitz, Stefano Massai, Don Marolf, and Thomas Van Riet for discussions and communications.  We thank Juan Maldacena and Eva Silverstein for crucial questions.

\end{document}